%% file: pscc2016_sccuc.tex
\documentclass[9pt]{IEEEtran4PSCC}

\ifCLASSINFOpdf
\else
\fi

\usepackage[cmex10]{amsmath}
\usepackage{graphicx,epstopdf,multirow,multicol,subfigure,booktabs,pgfplots}
\usepackage{amssymb,amsthm}
\usepackage{algorithm,algorithmic}
\usepackage[T1]{fontenc}
\usepackage[hidelinks]{hyperref}
\usepackage{booktabs}
\usepackage{cite}
\allowdisplaybreaks
\usepackage{amsfonts}

\begin{document}
\title{Unit Commitment with N-1 Security and Wind Uncertainty}

\author{
  \footnotesize
  \small
  Kaarthik Sundar$^\ddag$, Harsha Nagarajan$^{\dag}$,  Miles Lubin$^\P$, Line Roald$^\S$, Sidhant Misra$^{\dag}$, Russell Bent$^{\dag}$, Daniel Bienstock$^\$$ \\
$^\dag$ Los Alamos National Laboratory, Los Alamos, NM, United States. \\
$^\ddag$ Department of Mechanical Engineering, Texas A\&M University, College Station, TX, United States. \\
$^\P$ Operations Research Center, Massachusetts Institute of Technology, Cambridge, MA, United States. \\
$^\S$ Power Systems Laboratory, Department of Electrical Engineering, ETH Zurich, Switzerland. \\
$^\$$ Department of Industrial Engineering and Operations Research and \\Department of Applied Physics and Applied Mathematics, Columbia University, NY, United States.\\
}

\maketitle

\begin{abstract}
As renewable wind energy penetration rates continue to increase, 
one of the major challenges facing grid operators is the question of how to
control transmission grids in a reliable and a cost-efficient manner. 
The stochastic nature of wind forces an alteration of traditional methods for solving day-ahead and 
look-ahead unit commitment and dispatch. 
In particular, uncontrollable wind generation increases the risk of random component failures. 
To address these questions, we present an N-1 Security and Chance-Constrained Unit Commitment (SCCUC) that includes the modeling of generation reserves that respond to wind fluctuations and tertiary reserves to account for single component outages. The basic formulation is reformulated as a mixed-integer second-order cone problem to limit the probability of failure. We develop three different algorithms to solve the problem to optimality and present a detailed case study on the IEEE RTS-96 single area system. The case study assesses the economic impacts due to contingencies and various degrees of 
wind power penetration into the system and also corroborates the effectiveness of the algorithms.
\end{abstract}


\input{Intro}
\input{Formulation}

\input{Algo}
\input{NR}

\input{Conclusions}

\bibliographystyle{IEEEtran}
\bibliography{references.bib}

\end{document}

%% file: Intro.tex
\section{Introduction}
\label{Sec:intro}

Transmission grids play a vital role in the supply and the delivery of electric power. As renewable wind energy penetration rates continue to grow, reliable and cost-efficient operation of transmission grids becomes increasingly important. However, the stochastic nature of wind power necessitates an alteration of traditional methods for solving day-ahead and look-ahead Unit Commitment (UC) and generation dispatch. The fluctuations caused by uncontrolled wind generation can bring system components closer to their physical limits, making generator and line outages more likely to occur. As a result, power system operators are interested in securing the grid against component failures in the presence of these resources. 

More formally, the security of a power grid refers to its ability to survive contingencies, while avoiding disruption of service \cite{morison2005critical}. The failure to secure a power system could potentially result in cascading events \cite{andersson2005causes}. The concept of N-1 security assessment was developed to quantify this notion of security (see \cite{stott1987security} and references therein).  A power system is N-1 secure if it can survive all single component outages. While N-1 is an important security criteria, an N-1 secured power system still faces the risk of cascading events if one does not take into account the uncertain deviations of wind from its forecast value. 
One approach for controlling risk is chance constraints, where an upper bound on the probability of constraint violation is included in the OPF \cite{bienstock2014chance,vrakopoulou2013probabilistic}. 
Here, we adopt the latter approach and develop 
a comprehensive model that incorporates all aspects of day-ahead planning and security discussed above - UC for generators, N-1 security constraints for a line or a generator outage, chance constraints to ensure reliability with respect to wind in a stochastic sense, reserves from generators and tertiary (spinning) reserves.

\noindent
\textbf{Literature review:}
We review briefly some of the variations of the model explored in the literature and the algorithms that were proposed. In \cite{papavasiliou2013multiarea}, a stochastic UC variant was considered where the authors developed a two-stage dual-decomposition algorithm, accounting for wind via a scenario-based approach. The approach was validated on a 225 bus model of the California power system; the average computation time for a solving a 42-scenario model approximately 6 hours. A variation of this problem was studied in \cite{wang2012chance,margellos2013stochastic}, where the authors developed a sampling-based approach to account for wind uncertainty via a chance-constrained formulation (without \cite{wang2012chance} and with \cite{margellos2013stochastic} generation reserves modeling) and further corroborated their algorithms using Monte Carlo simulations against their deterministic variant on a modified IEEE 118-bus network and the IEEE 30-bus network respectively. Reference \cite{pandzic2015} developed a transmission-constrained UC formulation where the uncertainty is modeled using an interval formulation. They test their improved interval formulation against existing stochastic, interval and robust UC techniques in terms of solution robustness and cost on the IEEE RTS-96 test system. In contrast to this paper, none of these papers consider N-1 security constraints and the associated tertiary reserves modeling. 

Within the literature, the number of papers that do consider contingency modeling is limited. For instance, in \cite{vrakopoulou2013probabilistic}, authors solved a simpler OPF problem (without UC modeling) using a standard sampling-method to account for wind and N-1 security constraints. In terms of model comprehensiveness, the model most similar to ours is that of \cite{bertsimas2013adaptive}. They developed a two-stage adaptive robust UC model with security constraints and nodal net injection uncertainty. The model includes a deterministic uncertainty set, unlike the probability distribution of this paper, to model the wind. They proposed a Benders-based decomposition algorithm to handle line contingencies over a real-world large scale system, however they do not consider generator contingencies. It is the combination of both types of contingencies, especially the generator contingencies, that makes our problem more difficult to solve. Given the complexity of our problem resulting from the size of the network (24 buses) and N-1 security constraints on lines and generators for each one-hour time step over 24 hours, our algorithm compares favorably from a scalability perspective.



In this paper, we formulate the SCCUC as a large Mixed-Integer Second-Order Cone Program (MISOCP), leveraging the recent work in \cite{bienstock2014chance, lubin2015robust}. This formulation models two kinds of reserves, namely the generator reserves and the tertiary/spinning reserves to account for the wind fluctuations and single line or generator outage, respectively. We develop three approaches to solve the MISOCP with each approach building on its predecessor. The first algorithm is a traditional sequential linear outer approximation for the Second-Order Cone (SOC) constraints \cite{bienstock2014chance}. The second algorithm, the scenario-based decomposition, in addition to using the sequential outer approximation technique, exploits a block diagonal structure of the constraint matrix to decompose the formulation. The third algorithm is a Benders-like algorithm incorporating the classical Benders feasibility cuts to invalidate solutions to small SOC feasibility subproblems. Extensive computational experimentation based on the single area IEEE RTS-96 system is used to compare the effectiveness of each of the approaches. We also make a detailed comparison of the SCCUC formulation to the deterministic variant of the problem to illustrate the advantages of solving the SCCUC over its deterministic counterpart. 

%% file: Formulation.tex
\section{SCCUC Formulation \label{sec:formulation}}

Throughout this paper we utilize the linearized DC power flow model. While this model has its limitations, it is the model most generally used in the unit commitment literature. In order to quantify the primary complexities and benefits of this paper's core contribution (chance constrained response to wind fluctuations) we isolate this modification to the traditional unit commitment model. Future work will consider more realistic power flow models.

\subsection{Nomenclature \label{sec:nomenclature}}
\noindent \textbf{(i) Sets:}\\
$\mathcal B$ - set of buses, indexed by $b$ \\
$\mathcal L$ - set of lines, indexed by $\ell$ \\
$\mathcal G$ - set of all generators, indexed by $i$ \\
$\mathcal G_b$ - subset of generators  located at bus $b$ \\
$\mathcal S$ - set of generating units' start-up cost blocks, indexed by $s$ \\
$\mathcal K$ - set of generating units' production cost blocks, indexed by $k$\\
$\mathcal W$ - set of buses with wind farms \\
$\mathcal T$ - set of discretized times (hours), indexed by $t$ \\
$\mathcal C$ - set of contingencies, indexed by $c$ \\
\noindent \textbf{(ii) Binary decision variables:} \\
$x_i(t)$ - generator on-off status at time $t$ \\
$y_i(t)$ - generator 0-1 start-up status \\
$z_i(t)$ - generator 0-1 shut-down status \\
$w_{is}(t)$ - generator start-up block identification; $1$ if $i$ is started up at the beginning of hour $t$ after being down for $s$ hours, $0$ otherwise
\noindent \textbf{(iii) Continuous decision variables:} \\
$sc_i(t)$ - start-up cost of $i$ at hour $t$, \$\\
$p_i(t)$ - power output of $i$ for hour $t$, MW \\
$r^+_i(t)$ - generation up reserve power output of $i$ at hour $t$, MW\\
$r^-_i(t)$ - generation down reserve power output of $i$ at hour $t$, MW\\
$r^{up}_i(t)$ - tertiary reserve power output of $i$ at hour $t$, MW\\
$g^k_i(t)$ - power output on segment $k$ of cost curve of $i$ at $t$, MW\\
$\alpha_i(t)$ - participation factor of $i$ at hour $t$ \\
$f_{\ell}(t)$ - real power flow over line $\ell$ at hour $t$, MW \\
$\delta_i^c(t)$ - power provided by $i$ for contingency $c$ at hour $t$, MW\\
$\alpha^c_i(t)$ - participation factor of $i$ for contingency $c$ at hour $t$ \\
$f_{\ell}^c(t)$ - real power flow over line $\ell$ for contingency $c$ at hour $t$, MW \\
\noindent \textbf{(iv) Parameters:} \\
$\beta_{\ell}$ - susceptance of line $\ell$ \\
$p_i^{\min}$ - minimum output of generator $i$, MW \\
$p_i^{\max}$ - maximum output of generator $i$, MW \\
$p_{i,k}^{\max}$ - maximum power output of $i$ in production cost block $k$, MW\\ 
$d_b(t)$ - demand at bus $b$ for hour $t$, MW \\
$a^0_i$ - no-load cost of the generator $i$, \$ \\
$a^1_i$ - linear cost coefficient for $r_i^+$ and $r_i^-$ for $i$, \$/MW \\
$a^2_i$ - linear cost coefficient for $r_i^{up}$ for $i$, \$/MW \\
$K_i^k$ - slope of the $k^{th}$ segment of the cost curve for $i$, \$/MW \\
$f_{\ell}^{\max}$ - capacity of line $\ell$, MW \\
$\overline L_i$ - min. time $i$ has to run from the start of planning horizon, hrs\\
$\underline L_i$ - min. time $i$ has to be off from the start of planning horizon, hrs\\
$UT_i$ - minimum up-time of $i$, hrs \\
$DT_i$ - minimum down-time of $i$, hrs \\
$p^{\operatorname{up,init}}_i$ - time $i$ has been on before $t=0$, hrs \\
$p^{\operatorname{down,init}}_i$ - time $i$ has been on before $t=0$, hrs \\
$p^{\operatorname{on-off}}_i$ - on-off status of $i$ at $t=0$ ($1$ if $p^{\operatorname{up,init}}_i > 0$, $0$ otherwise) \\
$p_i(0)$ - power output of generator $i$ at $t=0$, MW\\
$RU_i$ - ramp-up limit of $i$, MW/hr\\
$RD_i$ - ramp-down limit of $i$, MW/hr\\
$c_{is}$ - cost of block $s$ of stepwise start-up cost function of $i$, \$\\
$\overline T_{is}$ - upper limit of block $s$ of the stepwise start-up cost of $i$, hrs\\
$\underline T_{is}$ - upper limit of block $s$ of the stepwise start-up cost of $i$, hrs\\
$\mu_b(t)$ - constant forecast output of wind farm at bus $b$ for hour $t$, MW\\
$\omega_b(t)$ - actual wind deviations from forecast $\mu_b(t)$, at hour $t$ \\
$r$ - index of the reference bus \\
$R$ - bounds on the amount of reserves that can be purchased, MW\\
$B$ - bus admittance matrix for the network \\
$B^c$ - bus admittance matrix for the network under contingency $c$

In the rest of the article, bold symbols denote random variables. In particular, $\boldsymbol \omega_b(t)$ is the random variable that models $\omega_b(t)$ for hour $t$. In the SCCUC, we assume that the deviations $\boldsymbol \omega_b(t)$ are independent and normally distributed with zero mean and variance $\sigma_b(t)^2$ (see \cite{bienstock2014chance}). We assume that $\boldsymbol \omega_b(t)$ are not correlated in time or across space geographically, however, this assumption can be relaxed to a certain degree w.l.o.g.
=These wind deviations drive the random fluctuations in the controllable generator injectors $\boldsymbol p_i(t)$, and line flows $\boldsymbol{f}_{\ell}(t)$. 


Finally, we let $\boldsymbol \Omega(t) = \sum_{b\in \mathcal W} \boldsymbol \omega(t)$ denote the total deviation in the wind from the forecast. For notional convenience, we use $\bar{p}(t)$, $\bar{\mu}(t)$, $\bar d(t)$, $\bar{\boldsymbol \omega}(t)$, $\bar{\delta}^c(t)$, $\bar\alpha(t)$, $\bar\alpha^c(t)$, $\bar r^+(t)$, $\bar r^-(t)$, $\bar r^{up}(t)$ to denote the vector of power generation, constant wind forecast, wind deviations, additional generation during contingencies, participation factors of the controllable generators, the participation factors during contingencies, generation up and down reserves and tertiary reserves respectively.
\subsection{Generation control \label{sec:gencontrol}}
As mentioned previously, we assume that the random wind deviations drive the controllable generator injections during each time period. Thus, the controllable generators respond proportionally to the wind fluctuations \cite{bienstock2014chance, lubin2015robust} as 
\begin{flalign}
\boldsymbol p_i(t) = p_i(t) - \alpha_i(t) \boldsymbol \Omega(t). \label{eq:genresponse}
\end{flalign}
Here, $\alpha_i(t) \geq 0$ is the participation factor for the controllable generator $i$. It was shown in \cite{bienstock2014chance} that when $\sum_i \alpha_i(t) = 1$, Eq. \eqref{eq:genresponse} guarantees balance of generation and load for every time period $t$. 
\subsection{Line flows \label{sec:lineflows}}
The random fluctuations in line flows $\boldsymbol f_{\ell}(t)$ for the line $\ell$ depend on the wind fluctuations implicitly, through the random bus angles $\boldsymbol \theta_{b}(t)$ which satisfy 
\begin{equation}
\bar p(t) + \bar{\mu}(t) - \bar d(t) + \bar{\boldsymbol \omega}(t) - \boldsymbol \Omega(t)\bar\alpha(t) = B\boldsymbol\theta(t).
\end{equation}
The bus admittance matrix $B$ is invertible after removing the row and column corresponding to the reference bus $r \in \mathcal{B}$. Following the DC power flow model\cite{wood2012power}, the flows $\boldsymbol f_{\ell}(t)$ are a linear function of the bus angles, hence we denote the $|\mathcal L| \times |\mathcal B|$ matrix $M$ as the linear map from power injections to line flows.
Then the random line flows for each line $\ell$ are computed as:
\begin{flalign}
\boldsymbol f_{\ell}(t) = M_{(\ell,\cdot)} \left(\bar p(t) + \bar{\mu}(t) - \bar d(t) + \bar{\boldsymbol \omega}(t) - \boldsymbol \Omega(t) \bar\alpha(t)\right). \label{eq:randomlineflow}
\end{flalign}  

\subsection{Reserve generation\label{sec:reserves}}
In the SCCUC, we model two types of reserves: generation reserves ($r^+_i(t), r^-_i(t)$) and the tertiary reserves ($r_i^{up}(t)$). The generation reserves are used to respond to wind fluctuations and the tertiary reserves are used to respond to generator outages. The linear cost coefficients for purchasing generation reserves and tertiary reserves from a generator $i$ at time is given by $a_i^1$ and $a_i^2$, respectively.

\subsection{Post-contingency generation outputs \label{sec:gout}}
The generation output of the generator $i$ after the outage of a generator $c$ during hour $t$ is modeled as
\begin{flalign}
\boldsymbol p^c_i(t) = p_i(t) - \alpha^c_i(t) \boldsymbol \Omega(t) + \delta^c_i(t) \label{eq:genresponsecontingency}
\end{flalign}
where, $\alpha^c_i(t) \geq 0$ is the new participation factor for the controllable generator $i$ corresponding to the outage of generator $c$ and $\boldsymbol p^c_i(t)$ is the random generator injection during the outage. To ensure power balance we enforce 
\begin{flalign}
\sum_i \alpha^c_i(t) = 1, \quad \sum_i \delta^c_i(t) = 0, \text{ and } \delta^c_c(t) = - p_c(t)  \label{eq:powerbalancegout}
\end{flalign}
For an outage of line $c$ during time $t$, the participation factors do not change \emph{i.e.,} $\alpha^c_i(t) = \alpha_i(t)$ and $\delta_i^c(t) = 0$ for all the controllable generators $i$. 

\subsection{Post-contingency line flows \label{sec:lout}}
The effect of a line or generator outage $c$ changes the topology of the system, which is represented by the matrix $M$ defined in Sec. \ref{sec:lineflows}. Let $M^c$ denote the matrix $M$ corresponding to the topology after an outage $c$. For generator outages, we have $M = M^c$. Using these notations, we model the line flow during a contingency $c$ as follows:
\begin{flalign}
\boldsymbol f^c_{\ell}(t) = M^c_{(\ell,\cdot)} \left(\bar p(t) + \bar \delta^c(t) + \bar{\mu}(t) - \bar d(t) + \bar{\boldsymbol \omega}(t) - \boldsymbol \Omega(t) \bar \alpha^c(t)\right) \label{eq:randomlineflowcontingency}
\end{flalign}
where, $\boldsymbol f^c_{\ell}(t)$ is the random line flow on the line $\ell$ during contingency $c$ at time $t$.

\subsection{Optimization problem\label{sec:optformulation}}
With the notations and modeling considerations in Sec. \ref{sec:nomenclature} -- \ref{sec:lout}, we present a formulation of the SCCUC. The objective function of the SCCUC minimizes the operating cost of the generators which includes the no-load cost, start-up cost, the running cost of all the generators and the cost of the generation and tertiary reserves, i.e.
\begin{equation}
\begin{split}
\min \sum_{i\in \mathcal G} \sum_{t \in \mathcal T} \bigl\{&a^0_i\cdot x_i(t) + \sum_{k\in \mathcal K} K_i^k \cdot g_i^k(t) + sc_i(t) +  \\
&\left[a^1_i\cdot (r^+_i(t)+ r^-_i(t)) + a^2_i\cdot r_i^{up}(t)\right] \bigr\} \label{eq:obj}
\end{split}
\end{equation}
The choice of the objective function is motivated by the models in \cite{pandzic2015,padhy2004,wang2008security}. The optimization is subject to constraints (\ref{eq:genresponse})-(\ref{eq:randomlineflowcontingency}) and the following constraints:
\subsubsection{Binary variable logic\label{sec:logic}}
\begin{flalign}
&y_i(t) - z_i(t) = x_i(t) - x_i(t-1) \quad \forall t\in \mathcal T, i\in \mathcal G, \label{eq:logic1}\\
&y_i(t) + z_i(t) \leq 1 \quad \forall t \in \mathcal T, i\in \mathcal G. \label{eq:logic2}
\end{flalign}
Constraint \eqref{eq:logic1} determines if the generator is started up or shut down at hour $t$ based of its on-off status between hour $t$ and $t-1$. Constraint \eqref{eq:logic2} ensures that a generator $i$ is not started up and shut down in the same hour $t$. 
\subsubsection{Generation limits\label{sec:genlimits}}
\begin{flalign}
&p_i^{\min} \cdot x_i(t) \leq p_i(t) \leq p_i^{\max} \cdot x_i(t) \quad \forall i \in \mathcal G, t\in \mathcal T, \label{eq:genlimits1} \\ 
&0 \leq r_i^-(t),r_i^+(t),r_i^{up}(t) \leq R \cdot x_i(t) \quad \forall i \in \mathcal G, t\in \mathcal T, \label{eq:genlimits2a} \\
&\delta_i^c(t) \leq r_i^{up}(t) \quad \forall i \in \mathcal G, t\in \mathcal T, c\in \mathcal C, \label{eq:genlimits2b} \\
&\sum_{n\in \mathcal G} r_n^{up}(t) \geq p_i(t) \quad \forall i\in \mathcal G, t\in \mathcal T, \label{eq:genlimits2c} \\
&p_i(t) - r_i^-(t) \geq p_i^{\min} \cdot x_i(t) \quad \forall i \in \mathcal G, t\in \mathcal T, \label{eq:genlimits3} \\
&p_i(t) + r_i^+(t) + r_i^{up}(t) \leq p_i^{\max} \cdot x_i(t) \quad \forall i \in \mathcal G, t\in \mathcal T, \label{eq:genlimits4} \\
&\operatorname{Pr}(r_i^-(t) \geq \boldsymbol \Omega(t) \alpha_i(t)) \geq 1-\varepsilon_i \quad \forall i \in \mathcal G, t\in \mathcal T, \label{eq:genlimits5a} \\
&\operatorname{Pr}(r_i^+(t) \geq -\boldsymbol \Omega(t) \alpha_i(t)) \geq 1-\varepsilon_i \quad \forall i \in \mathcal G, t\in \mathcal T, \label{eq:genlimits5b} \\
&\operatorname{Pr}(r_i^-(t) \geq \boldsymbol \Omega(t) \alpha_i^c(t)) \geq 1-\varepsilon_i \quad \forall i \in \mathcal G, t\in \mathcal T, c\in \mathcal C, \label{eq:genlimits6a} \\
&\operatorname{Pr}(r_i^+(t) \geq -\boldsymbol \Omega(t) \alpha_i^c(t)) \geq 1-\varepsilon_i \quad \forall i \in \mathcal G, t\in \mathcal T, c\in \mathcal C, \label{eq:genlimits6b} \\
&0 \leq \alpha_i(t), \alpha_i^c(t) \leq x_i(t) \quad \forall i \in \mathcal G, t\in \mathcal T, c\in \mathcal C. \label{eq:participation}
\end{flalign}
The constraints in \eqref{eq:genlimits1} -- \eqref{eq:genlimits4} enforce the generation limits and the reserve capacity limits for the generator $i$ at every hour $t$. In particular, constraint \eqref{eq:genlimits2c} ensures that the total tertiary reserves from all the generators during an hour $t$ must be greater than the maximum power generated by any generator during that hour. This guarantees that enough tertiary reserves are purchased at each hour to cover for any generator outage. The constraints in \eqref{eq:genlimits5a} -- \eqref{eq:genlimits6b} are the chance constraints on the generation limits. They ensure that generation reserves respond to wind fluctuations feasibly with high probability both during normal operation and contingencies. The constraints \eqref{eq:participation} impose the bounds on the participation factors.
\subsubsection{Piecewise linear production cost of the generators\label{sec:prodcost}}
\begin{flalign}
&p_i(t) = \sum_{k\in \mathcal K} g_i^k(t) \quad \forall i \in \mathcal G, t\in \mathcal T, \label{eq:production1} \\
&0 \leq g_i^k(t) \leq p_{i,k}^{\max} \cdot x_i(t) \quad \forall i \in \mathcal G, t\in \mathcal T, k \in \mathcal K. \label{eq:production2}
\end{flalign}
Constraint \eqref{eq:production1} defines the power generated by each generator $i$ and at each hour $t$ as the sum of power generated on each block of the production cost curve and the constraint \eqref{eq:production2} enforces the limits on the power generated on each block.
\subsubsection{Stepwise start-up cost of the generators\label{sec:startup}}
\begin{flalign}
&\sum_{s \in \mathcal S} w_{is}(t) = y_i(t) \quad \forall i \in \mathcal G, t\in \mathcal T, \label{eq:startup1}\\
&w_{is}(t) \leq \sum_{\underline T_{is}}^{\overline T_{is}} z_i(t-s) 
\quad \forall i \in \mathcal G, t\in \mathcal T, s\in \mathcal S, \label{eq:startup2}\\
&sc_i(t) = \sum_{s\in \mathcal S} w_{is}(t) \cdot c_{is} \quad \forall i \in \mathcal G, t\in \mathcal T.\label{eq:startup3} 
\end{flalign}
The start-up cost for a generator $i$ varies with the number of consecutive time periods $i$ has been off before it is started up. Constraint \eqref{eq:startup1} ensures exactly one start-up cost from the set of start-up cost blocks is chosen for the generator $i$. Constraint \eqref{eq:startup2} identifies the appropriate start-up block by implicitly counting the number of consecutive time periods the generator has been in the off state. Finally, constraint \eqref{eq:startup3} selects the actual start-up cost that shows up in the objective function.
\subsubsection{Minimum up and down time, ramping\label{sec:updownramp}}
\begin{flalign}
&x_i(t) =  p_i^{\operatorname{on-off}} \quad \forall i\in \mathcal G, t\leq \overline L_i + \underline L_i, \label{eq:updown1} \\
&\sum_{n=\overline{t}}^t y_i(n) \leq x_i(t) \quad \forall i\in \mathcal G, t\geq \overline L_i, \overline{t} = t-UT_i+1, \label{eq:updown2} \\
&\sum_{n=\underline{t}}^t z_i(n) \leq 1- x_i(t) \quad \forall i\in \mathcal G, t\geq \underline L_i, \underline{t} = t-DT_i+1, \label{eq:updown3} \\
&RD_i \geq p_i(t-1) - p_i(t) \quad \forall i \in \mathcal G, t\in \mathcal T, \label{eq:rampdown}\\
&RU_i \geq p_i(t) - p_i(t-1) \quad \forall i \in \mathcal G, t\in \mathcal T. \label{eq:rampup}
\end{flalign}
The constraint in Eq. \eqref{eq:updown1} sets the on-off status of the generator $i$ based on the initial conditions. Notice that both $\underline L_i$ and $\overline L_i$ will not take positive values simultaneously. The constraints in Eqs. \eqref{eq:updown2} and \eqref{eq:updown3} enforce the minimum up time and minimum down time constraints for generator $i$ for the remaining time intervals of the planning horizon. The constraints \eqref{eq:rampdown} and \eqref{eq:rampup} enforce the ramping limits on consecutive periods on every generator $i$.
\subsubsection{Power flow\label{sec:opf}}
\begin{flalign}
&\sum_{i\in \mathcal G} \alpha_i^c(t) = 1, \alpha_r^c(t) = 0  \quad \forall t \in \mathcal T, c \in \mathcal C,\label{eq:powerflow1b}\\ 
&\sum_{i\in \mathcal G} \alpha_i(t) = 1, \alpha_r(t) = 0  \quad \forall t \in \mathcal T, \label{eq:powerflow1a}\\ 
&\sum_{b\in \mathcal B} (p_b(t) + \mu_b(t) - d_b(t)) = 0 \quad \forall t\in \mathcal T, \label{eq:powerflow3a}\\
&\sum_{i\in \mathcal G} \delta^c_i(t) = 0 \text{ and } \delta^c_c(t) = - p_c(t) \quad \forall t \in \mathcal T, c \in \mathcal C, \label{eq:powerflow3b}\\
&\operatorname{Pr}(\boldsymbol f_{\ell}(t) \leq f^{\max}) > 1- \varepsilon_{\ell} \quad \forall \ell \in \mathcal L, t\in \mathcal T,  \label{eq:powerflow4a}\\
&\operatorname{Pr}(\boldsymbol f_{\ell}(t) \geq -f^{\max}) > 1- \varepsilon_{\ell} \quad \forall \ell \in \mathcal L, t\in \mathcal T,  \label{eq:powerflow4b}\\
&\operatorname{Pr}(\boldsymbol f^c_{\ell}(t) \leq f^{\max}) > 1- \varepsilon_{\ell}^c \quad \forall \ell \in \mathcal L, t\in \mathcal T, c\in \mathcal C, \label{eq:powerflow5a}\\
&\operatorname{Pr}(\boldsymbol f^c_{\ell}(t) \geq -f^{\max}) > 1- \varepsilon_{\ell}^c \quad \forall \ell \in \mathcal L, t\in \mathcal T, c \in \mathcal C. \label{eq:powerflow5b}
\end{flalign}
The power flow equations are adapted from \cite{bienstock2014chance, lubin2015robust} for multiple time periods. Constraints \eqref{eq:powerflow1b} -- \eqref{eq:powerflow1a} are the constraints on the participation factors that guarantees balance of generation and load (Sec. \ref{sec:gencontrol} and \ref{sec:gout}). Constraints \eqref{eq:powerflow3a} and \eqref{eq:powerflow3b} impose the demand-generation balance during normal operations and contingencies for all time periods. Finally, constraints \eqref{eq:powerflow4a} -- \eqref{eq:powerflow5b} are the chance constraints for the line flows during normal operations and during contingencies.

%% file: Algo.tex
\section{Algorithms} \label{sec:algo}
In this section, we present three different algorithms to solve the SCCUC problem. The common underlying ideas for all the three algorithms are: (i) we relax the formulation by ignoring a few constraints and provide the relaxed formulation to a branch-and-bound solver, (ii) whenever a feasible solution is obtained to this relaxed problem, we check if the solution is feasible for the constraints that were ignored, (iii) if one or more constraints are violated for the current feasible solution, then we add a cut to the original problem that invalidates the solution and continue solving the problem. Each algorithm presented in this section differs in steps (i) and (iii). All algorithms are implemented within a single branch-and-bound tree by using solver callbacks. 

\subsection{Outer approximation\label{sec:oa}}
The formulation in Sec. \ref{sec:formulation} contains chance constraints (Eq. \eqref{eq:powerflow4a}--\eqref{eq:powerflow5b}) that can be reformulated as Second-Order Cone (SOC) constraints. The chance constraints in Eq. \eqref{eq:genlimits5a}--\eqref{eq:genlimits6b} are a special case where the reformulation is linear (see \cite{lubin2015robust}). While this reformulation of chance constraints as SOC constraints is useful, \cite{bienstock2014chance} also observed that off-the-shelf commercial solvers were not able to handle large scale CCOPF instances i.e., continuous SOC problems. 
So, to address this issue, we use the following approach: we omit the reformulated SOC constraints corresponding to Eq. \eqref{eq:powerflow4a}--\eqref{eq:powerflow5b} from the formulation we provide to the solver. Whenever the solver obtains an integer feasible solution to this relaxed problem, we check if it satisfies all the SOC constraints that were ignored. If not, we add a linear outer approximation of the infeasible SOC constraint and continue solving the original problem. This process of adding linear outer approximations of violated SOC constraints sequentially has been observed to be computationally efficient for the  CCOPF and robust CCOPF problems \cite{bienstock2014chance, lubin2015robust}. 

We provided the formulation, as stated in Sec. \ref{sec:optformulation}, to the modeling tool JuMPChance \cite{JuMP}, which enables the user to select between solutions via sequential outer approximation or via reformulation to an SOC problem. Despite using sequential outer approximations to address the issue of the SOC constraints, solving the full problem, as is, can be time consuming even for small SCCUC instances. We also observe that the constraints of the SCCUC formulation have an inherent block-diagonal structure with a few coupling constraints that can be exploited to develop more efficient exact algorithms. In the following section, we discuss a modified version of a traditional scenario-based decomposition algorithm that exploits this block diagonal structure of the constraint matrix to compute optimal solutions to the SCCUC.

\subsection{Scenario-based decomposition\label{sec:sbd}}
In this section, we present a scenario-based decomposition (SBD) approach to solve the SCCUC. This algorithm is an improvement to the outer approximation algorithm. We handle the SOC constraints in exactly the same way as for the outer approximation. In addition, we also leave out constraints corresponding to a subset of contingencies $\mathcal C_1$ and solve the relaxed problem. Whenever the solver obtains a feasible solution to the relaxed problem, in addition to checking if all the SOC constraints are satisfied by the current feasible dispatch, we also check if the dispatch violates any of the contingencies in $\mathcal C_1$. The violated SOC constraints are added as linear outer approximation cuts and the constraints corresponding to the infeasible contingencies in set $\mathcal C_1$ are directly added to the relaxed problem. We note that once all the constraints corresponding to an infeasible contingency in $c_1 \in \mathcal C_1$ are added to the relaxed problem, this contingency will not be violated by any subsequent feasible solutions and hence, $c_1$ can be removed from the list of contingencies $\mathcal C_1$. Checking if the contingencies in $\mathcal C_1$ are feasible for the current dispatch involves solving an SOC feasibility problem for every hour $t$ and every contingency $c_1 \in \mathcal C_1$, given by the constraints \eqref{eq:genlimits2b}, \eqref{eq:genlimits6a}, \eqref{eq:genlimits6b}, \eqref{eq:participation}, \eqref{eq:powerflow1b}, \eqref{eq:powerflow3b}, \eqref{eq:powerflow5a}, and \eqref{eq:powerflow5b}. These are small SOC feasibility problems and can potentially be solved in parallel. This small improvement to the outer approximation algorithm results in a speed up of a factor of two as we will observe in Sec. \ref{sec:case}. 

\subsection{Benders decomposition\label{sec:benders}}

The SBD approach is effective when only a few of the contingencies are ``active'' in the final solution. However, we found that in a number of cases, especially when we decrease the allowed probability violations $\epsilon_\ell^c$, a large fraction of the contingencies needed to be added before convergence is achieved. As an alternative, we developed a Benders-like decomposition. The relaxed problem we provide to the solver is the entire formulation defined in Section~\ref{sec:formulation} except for the constraints~\eqref{eq:powerflow5a}-\eqref{eq:powerflow5b}. Instead of treating these constraints as is, we use an extended formulation which avoids forming (and storing in memory) the entire dense matrix $M^c$ defined in Section~\ref{sec:lout} for each contingency. Benders decomposition (generalized to SOCP subproblems) is then applied to this extended formulation. Whenever an integer feasible solution is found by the solver, we invoke the Benders cut generation procedure briefly described below.

For a contingency $c \in \mathcal{C}$, define $LF_{c}(\bar p(t),\bar \delta^c(t),\bar \alpha^c(t))$ as the set of $(\theta^c,\gamma^c,f^c) \in \mathbb{R}^{2|\mathcal{B}|+|\mathcal{L}|}$ satisfying $\gamma_r^c = \theta_r^c = 0$,

\begin{flalign}
&\sum_{n\in \mathcal B} B^c_{bn}\theta_n^c = \sum_{i\in \mathcal G_b} \left[p_i(t) + \delta_i^c(t)\right] + \mu_b(t) - d_b(t) \quad \forall b\in \mathcal B, \\
&\sum_{n\in \mathcal B, n\neq r} B_{bn} \gamma_n^c = \sum_{i \in \mathcal G_b} \alpha_i^c(t) \quad \forall b\in \mathcal B\setminus \{r\}, \\
&f^c_{mn} = \beta_{mn}(\theta^c_m - \theta^c_n) \quad \forall (m,n) \in \mathcal L^c,
\end{flalign}
\begin{equation}
\begin{split}
&\operatorname{Pr}(f^c_{mn} + \beta_{mn} \boldsymbol \Omega(t) (\gamma_n^c - \gamma_m^c) + \beta_{mn} \bar{\boldsymbol \omega}^T(t)(\pi^c_m - \pi^c_n) \\
& \leq f_{mn}^{\max}) \geq 1-\epsilon^c_{mn}
\quad\forall (m,n) \in \mathcal L^c,
\end{split}
\end{equation}
\begin{equation}
\begin{split}
&\operatorname{Pr}(f^c_{mn} + \beta_{mn} \boldsymbol \Omega(t) (\gamma_n^c - \gamma_m^c) + \beta_{mn} \bar{\boldsymbol \omega}^T(t)(\pi^c_m - \pi^c_n) \\
& \geq -f_{mn}^{\max}) \geq 1-\epsilon^c_{mn}
\quad \forall (m,n) \in \mathcal L^c,
\end{split}
\end{equation}
where, $\pi^c_b$ is the $b$th row of the inverse of the admittance matrix $B^c$, after excluding the row and column corresponding to the reference bus $r$. The set $\mathcal L^c$ is defined as $\mathcal L \setminus \{c\}$ if $c$ is a line contingency and $\mathcal L$ for generator contingencies.

By~\cite{lubin2015robust,bienstock2014chance}, the constraints~\eqref{eq:powerflow5a}-\eqref{eq:powerflow5b} are satisfied if and only if the set $LF_{c}(\bar p(t),\bar \delta^c(t),\bar \alpha^c(t))$ is not empty. We therefore test feasibility of the solution which the solver finds by solving the SOCP feasibility problem corresponding to $LF_{c}(\bar p(t),\bar \delta^c(t),\bar \alpha^c(t))$ for each contingency and time period. If the problem is infeasible, we compute, via SOCP duality, a cut analogous to the Benders feasibility cut which invalidates the solution $(\bar p(t),\bar \delta^c(t),\bar \alpha^c(t))$~\cite[Prop. 2.4.2]{BTNemirovskiLectures}. These cuts are added by using solver callbacks within the branch-and-bound tree as previously discussed. A technical challenge we encountered was obtaining valid dual rays to infeasible SOCPs, a feature not supported by CPLEX~\cite{cplex}. Instead, we used Mosek~\cite{mosek}, which has this functionality, to solve the SOCP subproblems.

%% file: NR.tex
\section{Case Study}
\label{sec:case}


In this section, we demonstrate the benefits of solving the SCCUC relative to the deterministic version of the unit commitment problem using the single area IEEE RTS-96 system \cite{RTS96}. The comparison is based on a variety of factors including nominal operational cost, number of line and generator violations, amount of generation and tertiary reserves allotted, etc. We also investigate the performance of proposed algorithms with regards to computation time and scalability.


\subsection{Test system and wind data\label{sec:testdata}}
We use the IEEE single-area RTS-96 system with modifications performed on the base system similar to the ones described in \cite{pandzic2015}. The system comprises of 24 buses including 17 load buses, 38 transmission lines and 32 conventional generators. The total installed capacity of the generators is 3405 MW. Among the 32 generators, 2 are nuclear and 1 is a hydro generator. The NREL Western Wind dataset \cite{potter2008creating} provides the wind data. Wind farms locations are mapped to the IEEE RTS-96 respecting the lengths of the lines (see \cite{pandzic2015}). The test system contains a total of 9 wind farms with a total generation capacity of 3900 MW. The locations of the wind farms, the individual generation capacity of each wind farm; the stepwise generation cost, start-up cost, ramping restrictions, up and down-time restrictions for each generator; and the load profile data for a 24-hour period used for the case study are made available by authors in \cite{pandzic2015} and \cite{data}. Using the 1000 wind generation scenarios for each wind farm generated in \cite{pandzic2015} via various statistical methods, the mean and standard deviation of the wind power injection for each time period was estimated assuming that the wind power injections for each hour are independent normally distributed random variables. The cost coefficients $a_i^0,~a_i^1,$ and $a_i^2$ for the generation and tertiary reserves, respectively, are adopted directly from the IEEE RTS-96 generation cost coefficient data and the wind power is assumed to have zero marginal cost.




\subsection{Performance of proposed algorithms}
The SCCUC formulation has four user-defined parameters namely $\epsilon_i,~\epsilon_i^c,~\epsilon_{\ell},$ and $\epsilon^c_{\ell}$. For the rest of the computational experiments in the paper, we set their values to $1\%,~2\%,~10\%$, and $20\%$ respectively. We then sequentially vary the loading levels and the wind penetration levels and solve the resulting SCCUC instances using each of the algorithms proposed in Sec. \ref{sec:algo}. The first algorithm is the sequential linear outer approximation algorithm (OA), the second is the SBD presented in Sec. \ref{sec:sbd}, and the last one is the Benders decomposition in Sec. \ref{sec:benders}. For the SBD, the set $\mathcal C_1$ is the set of all generator contingencies. The computations were carried out on a Dell Precision T5500 workstation (Intel Xeon E5630 processor \@ 2.53GHz, 12GB RAM). For all the runs, the optimality tolerance was set to 1\%. The algorithms were implemented using Julia and JuMPChance \cite{julia} with CPLEX and Mosek \cite{cplex, mosek} as the LP and conic solvers respectively. Table \ref{tab:times} shows the computation time of the three algorithms for varying wind penetration ($W\%$) and load levels ($L\%$). We observe from the results in Table \ref{tab:times} that the Benders decomposition outperforms the other two algorithms for all the test instances. We also note that this trend was observed consistently for different choices of subsets of of contingencies, $\mathcal C_1$, in the SBD. Hence, throughout the rest of the article, we use the Benders decomposition algorithm for all the runs.

\begin{table}
\caption{Computation times in seconds. }
\footnotesize
\centering
\resizebox{0.5\textwidth}{!}{%
\begin{tabular}{crrrrrrrrr}
\toprule
 & \multicolumn{3}{c}{$W\%=10\%$} & \multicolumn{3}{c}{$W\%=20\%$} & \multicolumn{3}{c}{$W\%=30\%$} \\
 \cmidrule(lr){2-4}
 \cmidrule(lr){5-7}
 \cmidrule(lr){8-10} 
 $L\%$ & OA & SBD & Benders & OA & SBD & Benders & OA & SBD & Benders \\
 \midrule
 70 & 1483 & 586 & 58& 507 & 467 & 110& 1230 & 467 & 54\\
 80 & 1467 & 743 & 63 & 1262 & 580 & 130& 1641 & 571 & 115\\
 90 & 1046 & 691 & 106 & 951 & 803 & 39& 1736 & 552 & 121\\ 
 100 & 1117 & 713 & 94 & 1353 & 720 & 85& 984 & 716 & 112\\
 \bottomrule
\end{tabular}}
\label{tab:times}
\end{table}

\subsection{Comparison of the SCCUC to its deterministic counterpart}
We now compare the performance of the SCCUC with its deterministic counterpart, which assumes $\boldsymbol \omega_b(t)=0$. 
Both the deterministic and chance constrained unit commitment with N-1 security constraints are solved for a case where the forecast wind power production accounts for 20\% of the total load. Since the deterministic unit commitment with N-1 security constraints assumes there are no wind fluctuations, it will not require any generation reserves to cover for the wind power fluctuations. In order to make a fair comparison, we assume that the system operator maintains a minimum generation reserve requirement equal to 0.5\% of the load and impose this constraint as a part of the optimization problem. Furthermore, to obtain a more interesting case, the transmission capacities of the IEEE RTS-96 system were decreased to 90\% of their original base case value in \cite{pandzic2015}. 

\begin{figure}[!t]
\includegraphics[width=0.95\columnwidth]{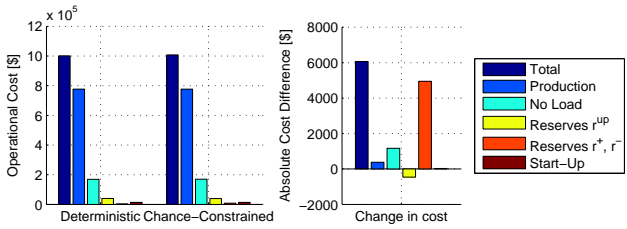}
\centering
\caption{Total cost and value of the different cost components for the deterministic and chance constrained unit commitment (left). The cost difference (right) shows the change in cost when moving from the deterministic to the chance constrained unit commitment.}
\label{fig:costs}
\end{figure}

The total cost of the unit commitment and the different cost components are shown in Fig. \ref{fig:costs}, with the deterministic and chance constrained cost on left and the cost differences to the right. The number of committed units and the amount of allocated reserves are shown in Table \ref{units_reserves}. We observe that the total cost of the SCCUC solution is only slightly greater than the cost of its deterministic counterpart, with a major difference showing up in the cost of generation reserves and a minor one in the no-load cost. The reason for the increased cost of generation reserves in the chance-constrained version is that it allocates twice as much generation reserves as the deterministic case in order to accommodate the wind power fluctuations; as mentioned previously the deterministic counterpart is immune to wind fluctuations and it allocates the bare minimum amount of generation reserves as required by the system operator (0.5\% of the load). The same reason is valid for the higher number of committed units for the chance-constrained case and the increased no-load cost (see Table \ref{units_reserves}).

\begin{table}
\caption{Number of committed units and allocated reserves}
\label{units_reserves}
\centering
\begin{tabular}{l c c c }
\toprule
                   & Committed  & Tertiary      & Generation  \\
                   & Units & Reserves [MW] & Reserves [MW]  \\
\midrule
Deterministic      & 417        & 5383.7        & 482.1\\
Chance constrained & 421        & 5357.5        & 954.3\\
\bottomrule
\end{tabular}
\end{table}

Next, we compare the two variants in their ability to handle wind fluctuations during normal operations and contingencies. To do so, we solve both the variants and obtain the corresponding optimal solution. To assess the performance of both the deterministic and the chance constrained solutions under simulated wind power conditions, we evaluate the line flows and generator outputs using the optimal dispatches for 1000 wind realizations and compute the number of generator and line violations that occur for each of the realization. The samples are drawn from the multivariate normal distribution whose means and variances were known a priori and were used as input to the SCCUC i.e., we assume perfect knowledge of the probability distribution of the wind fluctuations.

\begin{figure}[!t]
\includegraphics[width=0.9\columnwidth]{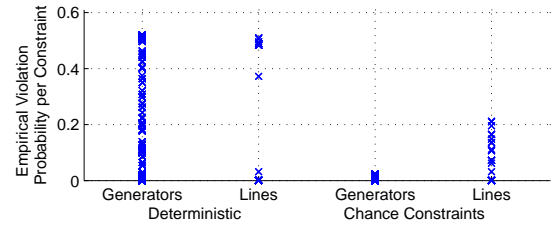}
\centering
\caption{Empirical violation probability of each generation and line constraint in the deterministic (left) and the chance constrained (right) solution.}
\label{fig:EmpViol}
\end{figure}

The empirical violation probabilities, evaluated for each constraint separately, are shown in Fig. \ref{fig:EmpViol}. We observe that the maximum value of the empirical violation probabilities for the optimal dispatch of the deterministic problem is 50\% for both the generator and line limit constraints. Interestingly, the optimal solution of the chance-constrained problem is able to effectively limit the empirical violation probabilities to 2\% and 20\% for generators and lines, respectively. We note that these violations include the ones that occur during single component outages. 

\begin{figure}[!t]
\includegraphics[width=0.95\columnwidth]{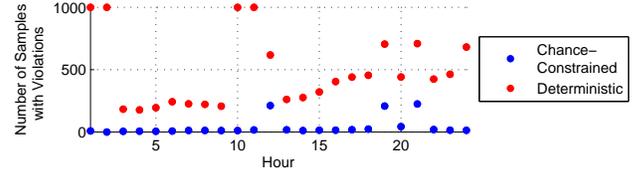}
\centering
\caption{The number of wind fluctuation samples (out of 1000) that lead to one or more constraint violations.}
\label{fig:SampleViol}
\end{figure}

For a system operator, it is tantamount not only to limit the violation probability for a specific line or generator limit constraint, but also to reduce the frequency at which any \emph{any} line or generator limit is violated during a 24-hour period. Hence, Fig. \ref{fig:SampleViol} compares the number of wind realizations out of 1000 that actually lead to at least one constraint violation during each hour of the day.  It is clear from the Fig. \ref{fig:SampleViol} that for the optimal dispatch obtained from the deterministic problem, the empirical probability that at least one constraint is violated during each hour varies from 20 to 100\%, while for the optimal solution to the chance constrained version of the problem, the risk is much lower, with a meager 10-20 samples creating violations for most hours of the day and with 225 being the maximum number of samples with violations during any hour of the day. 
%

\subsection{Influence of wind penetration and wind variability}
To assess the influence of wind power penetration and wind power variability on the cost of unit commitment and amount of reserves, we generate two classes of test instances. In the first class of instances, we fix the wind penetration level and vary the normalized standard deviation of the wind fluctuation linearly and in the second class, we do the vice-versa. The Table \ref{tab:windlevels} shows the variation of the number of committed units, the amount of generation reserves and the total SCCUC cost for different levels of wind penetration. We observe that the increase in the wind penetration levels decrease the overall SCCUC cost. The amount of generation reserves is more or less constant because the normalized standard deviation of the wind fluctuations is a constant for the first class of test instances. The decrease in the number of units committed can also be explained by the fact that power production from the conventional generators in the system decreases with increasing levels of wind penetration.

\begin{table}[htbp]
\caption{Number of committed units, generation reserves (MW), and SCCUC cost (\$) for varying levels of wind penetration}
\footnotesize
\centering
\begin{tabular}{cccc}
\toprule
 Wind  & Committed & Generation & Total SCCUC \\
 penetration [\%] & Units & Reserves [MW] & Cost [\$] \\
 \midrule
 5  & 469   & 284.47 & 1177691.41 \\
 10 & 443   & 292.04 & 1111250.49 \\
 15 & 431   & 292.64 & 1050269.73 \\
 20 & 415   & 290.09 & 996352.13 \\
 25 & 395   & 284.09 & 966712.09 \\
 \bottomrule
\end{tabular}
\label{tab:windlevels}
\end{table}

The results in Table \ref{tab:windvariance} indicates that the number of units committed remains a constant as the variation in wind power output is increased. One might expect the number of committed units to increase similar to the previous study, but instead the optimization problem chooses to increase the amount of generation reserves being produces from the already committed units. This reveals a trade-off between the cost of the generation reserves and the no-load cost. For the test instances chosen, the no-load cost was observed to be slightly greater than generation reserve cost and hence the behaviour results. The same reasoning holds for the cost of the generation reserves.

\begin{table}[htbp]
\caption{Number of committed units, generation reserves (MW), and generation reserve cost (\$) for varying levels of normalized standard deviation}
\footnotesize
\centering
\begin{tabular}{cccc}
\toprule
 Normalized $\sigma$  [\%] & Committed & Generation & Total Gen. \\
  & Units & Reserves [MW] &  Reserves Cost [\$] \\
 \midrule
 5  & 414   & 305.58 & 1830.12 \\
 10 & 414   & 321.39 & 1928.40 \\
 15 & 414   & 334.80 & 2005.43 \\
 20 & 413   & 349.41 & 2093.09 \\
 25 & 414  & 364.01& 2180.74 \\
 \bottomrule
\end{tabular}
\label{tab:windvariance}
\end{table}


%% file: Conclusions.tex
\section{Conclusions}
\label{sec:conc}

In this paper, we have presented a MISOCP formulation for the SCCUC problem in the presence of wind fluctuations. To the best of our knowledge, this is the first UC formulation in the literature that includes N-1 security constraints on lines and generators, wind fluctuations, and generator and tertiary reserves. Three algorithms were developed to compute an optimal solution to the SCCUC. The effectiveness of the proposed approach and its advantages over its deterministic counterpart was demonstrated through extensive computational experiments on the single area IEEE RTS-96 system. The results indicate that the proposed formulation is effective and can result in better technical performance including fewer violations of transmission line limits and generator limits during normal operations and during single line or generator outages when compared to its deterministic counterpart. Future work includes (i)  generalizations to account for errors in estimating the parameters of the probability distributions for the wind fluctuations, (ii) developing a computationally tractable method to handle correlated wind fluctuations, and (iii) extension of the SCCUC formulation to consider more realistic power flow models. 

\section*{Acknowledgements} This work was supported by the Advanced Grid Modeling Program of the Office of Electricity within the U.S Department of Energy and the Center for Nonlinear Studies at Los Alamos National Laboratory.